\documentclass[10pt, conference, compsocconf]{IEEEtran}
\pdfoutput=1
\usepackage{amsmath}
\usepackage{amsfonts}
\usepackage{setspace}
\usepackage{array}
\usepackage{amssymb}
\usepackage{graphicx}
\usepackage{url}
\usepackage{multicol}
\usepackage{stfloats}
\usepackage{subfigure}
\usepackage[usenames]{color}
\usepackage{flushend}
\usepackage{booktabs}
\usepackage{changes}
\usepackage{eurosym}
\usepackage{rotating}

\newcommand{\bi}{\begin{itemize}}
\newcommand{\ei}{\end{itemize}}
\newcommand {\beq}{\begin{equation}}
\newcommand {\eeq}{\end{equation}}
\newcommand {\be}{\begin{enumerate}}
\newcommand {\ee}{\end{enumerate}}

\begin{document}

\title{Analysis of publicly disclosed information in Facebook profiles}

\author{\IEEEauthorblockN{Reza Farahbakhsh\IEEEauthorrefmark{1},
Xiao Han\IEEEauthorrefmark{1},
\'Angel Cuevas\IEEEauthorrefmark{1}\IEEEauthorrefmark{2} and
No\"{e}l Crespi\IEEEauthorrefmark{1} }

\IEEEauthorblockA{\IEEEauthorrefmark{1}Institut Mines-T\'el\'ecom, T\'el\'ecom SudParis\\
\{reza.farahbakhsh, han.xiao, noel.crespi\}@it-sudparis.eu}
\IEEEauthorrefmark{2}Universidad Carlos III de Madrid\\
{acrumin@it.uc3m.es}\\
}


\maketitle

\begin{abstract}
%

Facebook, the most popular Online social network is a virtual environment where users share information and are in contact with friends. Apart from many useful aspects, there is a large amount of personal and sensitive information publicly available that is accessible to external entities/users. In this paper we study the public exposure of Facebook profile attributes to understand what type of attributes are considered more sensitive by Facebook users in terms of privacy, and thus are rarely disclosed, and which attributes are available in most Facebook profiles. Furthermore, we also analyze the public exposure of Facebook users by accounting the number of attributes that users make publicly available on average. To complete our analysis we have crawled the profile information of 479K randomly selected Facebook users. Finally, in order to demonstrate the utility of the publicly available information in Facebook profiles we show in this paper three case studies. The first one carries out a gender-based analysis to understand whether men or women share more or less information. The second case study depicts the age distribution of Facebook users. The last case study uses data inferred from Facebook profiles to map the distribution of worldwide population across cities according to its size.
\end{abstract}


\section{Introduction}
\label{sec:Introduction}

Facebook is the most popular On-line Social Network (OSNs) with more than one billion subscribers. Users mainly utilize Facebook to share their opinions, interests, personal content like pictures with users who are connected to them. An important element that Facebook incorporates is the possibility of defining detailed profile where users provide information about themselves. In Facebook we find more than 20 different attributes that can be utilized in a user profile. Those attributes include potentially sensitive information such as contact info, birth date, current city, home town, employers, college, high school, etc. Furthermore, together with personal details, Facebook users can complete their profiles by expressing their interests in different categories such as music, movies, books, etc., 
which in many cases facilitates deriving sensitive information from a user (e.g. personality characteristics, political leanings). Depending on the person, their status and this information's social context, publicly disclosing this sort of information could lead to some serious privacy issues. To avoid or at least mitigate these problems, Facebook allows each user to define a degree of privacy for different attributes in the profile. That is, for each attribute, a Facebook user can decide among several privacy options: $(i)$ leaving an attribute blank so that no one will get access to that information; $(ii)$ filling out an attribute and defining its privacy level as \textit{``only me''} meaning only the user has access to that information; $(iii)$ defining the attribute privacy level as \textit{``friends''}
; $(iv)$ defining the attribute privacy level as \textit{``friends of friends''}
; $(v)$ defining the attribute privacy level as \textit{``custom''} 
and $(vi)$ defining the privacy level as \textit{``public''} so that any user can access that information. Base on the Facebook strategies by default most of the attributes are publicly available except the birthday, Political views, Religion and Contact Info that are in the level of ``only Friends". For these attributes users can change the privacy level to public or more private.

The information included in the profile of Facebook users is precious for external users/entities and these have very divergent objectives, from non-lucrative activities such as research to lucrative ones, including marketing campaigns. Given the privacy management provided by Facebook, external entities can only access attributes that have been defined as ``public'' by users. Therefore, an important question to answer is what is the amount of public information that an external user/entity can find in Facebook profiles. In other words, what is the portion of Facebook users that publicly disclose (i.e. indicate privacy level ``public'') each of the profile attributes. By answering this question for each attributes we will be able to understand which type of information is considered more sensitive by Facebook users, and to the contrary, what are the attributes experiencing major public exposure.

Toward this end we have collected the public profiles of 479K randomly-selected Facebook users, and analyze 19 of the profile's attributes by computing the portion of the collected users that publicly disclose each attribute in their profiles. We divide the analyzed attributes into two groups: \textit{personal} and \textit{interest-based} attributes. The former category refers to attributes that contain personal life information about the user (e.g. location, education,  work history, etc). Interest-based attributes, on the other hand reflect the tastes of Facebook users, revealed by their preferences (e.g. in music, television, sport teams, etc). The results will let us determine the attributes that users consider more sensitive. Furthermore, we explore the correlation degree among the different personal attributes. That is, determining if a user disclosing a personal attribute $A$ has some relation to that user also publicly sharing a different attribute $B$. In order to get a meaningful answer, in this paper we correlate 9 personal attributes 2 by 2.

Our attribute-based analysis tells us how much information can be retrieved for a particular attribute, but it does not contribute anything regarding the expected amount of information that we can extract from a typical Facebook user. Therefore,  we seek to understand the public exposure habits of Facebook users themselves. To that end we have defined a very simple yet meaningful metric that accounts for the number of attributes that are publicly disclosed in a Facebook profile, and refer to it as the \textit{Degree of Public Exposure (DPE)}. The DPE ranges from 0 for user profiles that do not have any attribute publicly available, to 19 when a user has made all the analyzed attributes available, including personal and interest-based attributes. Hence, we can assign each of the 479K users in our dataset a DPE value. Using this metric and our dataset we are able to 
identify what type(s) of users present a higher degree of public exposure.

Finally, in the last part of this paper,  we define three simple use cases to illustrate how some external entities can utilize the information that is publicly accessible in Facebook. First, we perform a gender-based division of different personal attributes to discover whether men or women show a significant predisposition to publicly disclose particular type of information. Second, we depict the distribution of the ages of our 479K  Facebook users based on those users that publicly share their ages. Third, we check the accuracy that could be achieved by using Facebook users as an estimator for the distribution of the world wide population in cities.

The main observations extracted from the paper are:

\noindent $(i)$ Friend-list is the attribute with the largest public exposure with almost 63\% of users publicly sharing their contacts, whereas a users' age (i.e. Birth date attribute) rate as having the highest privacy value for from Facebook users, since only 3\% disclose this information.

\noindent $(ii)$ There are strong correlations between \textit{Current City} and \textit{Home Town} attributes. This may be because both attributes provide a type of ``location'' information, and users revealing one tend to also share the other. In addition, we found a second high correlation between education (i.e. \textit{College}) and professional experience (i.e. \textit{Employers}) attributes. 

\noindent $(iii)$ The average Facebook user makes more than four attributes publicly available in their profiles.

\noindent $(iv)$ Men show a larger public exposure than women for all personal attributes except \textit{birth date}. This exception is very surprising given the widespread assumption that women tend to hide their real age more than men.

\noindent $(v)$ The age range most-represented, based on the publicly available information, is 18-25. That range accounts for 1/2 of the users among those making their birth date publicly available. 

\noindent $(vi)$ We show that Facebook data very accurately estimates the portion of people that live in cities of more than 5 million (according to a recent United Nation report \cite{UN_City}). It also provides an accurate estimation for the proportion of people living in cities ranging between 500K-1M inhabitants, whereas it has a 10\% deviation for cities of less than 500K and for cities with between 1M and 5M citizens.

The rest of this paper is organized as follows: Section \ref{sec:related_work} presents related work and section \ref{sec:methodology} describes our data collection techniques as well as the attributes definition. Section \ref{sec:exposure_attributes} and \ref{sec:exposure_users} discuss the disclosure degree of the Facebook profiles attributes and section \ref{sec:Group_analysis} shows the usage of profile's disclosed information by analyzing three attributes of the profiles. Finally we conclude the paper in Section \ref{sec:Conclusion}.

\section{Related work}
\label{sec:related_work}

We explore the prior efforts regarding to user privacy in online social networks that establish the basis for our work.
%
%
In a concept similar to our study, Quercia et al. (2012) \cite{Quercia_Facebook_Privacy} found
a correlation with the degree of openness and gender, using a dataset of 1323 profiles from the United States. Our work has many distinctions from this study. Firstly, our dataset is much larger and broader (479K profiles widely distributed throughout the world compared to a little more than 1K profiles exclusively from U.S.). Secondly, our data was gathered directly from Facebook profiles, while Quercia et al. used a form of questionnaire administered by a specific Facebook application. Lastly, we study most of the available attributes in the FB profiles, and for some of them we deeply investigated the correlation between the attribute type and profile characteristics. They also concluded that men tend to make their profile information more publicly available.  In another work by these authors \cite{Quercia_popular_facebook}, they study the personality characteristics of popular Facebook users. 

Gross et al. in \cite{privacy-facebook-gross} studied the patterns of information revelation in Facebook. They analyzed just around 4K Carnegie Mellon University students' profiles. 
They evaluate the amount of information students disclose and their usage of the site's privacy settings. 

In other work, Chang et al. \cite{ChangRBM10} studied the privacy attitudes of U.S. Facebook users of different ethnicities. Another U.S.-based study \cite{Lewis_thetaste} used a questionnaire and with considering 1,710 students' profiles shows that women are more likely to maintain a higher degree of profile privacy than men; and that having a private profile is associated with a higher level of online activity. The authors in \cite{All_about_me} examined disclosure in Facebook profiles looking at only 400 Facebook profiles. 

In a study of the Facebook users' profile attributes, authors in \cite{leakage_infocom} present  a method to estimate the birth year of 1M Facebook users in New York City, based on the information available on their profiles, such as their friends. Authors in \cite{You_are_who_you_know} examined the possibility of using the attributes of users, in combination with their social network graph, to predict the attributes of another user in the network. Other similar work \cite{lampe} presents a study of Facebook profile attributes by analyzing a dataset of 30,773 Facebook profiles. They were able to determine which profile attributes are most likely to predict friendship links. They explore how profile attributes relate to the \#Friends of a user's profile.
An investigation of Facebook users' privacy evolution in a dataset of a large sample of New York City (NYC) Facebook users, was presented in \cite{percom_Facebook}. That study shows how the close/disclose status of profiles attributes changed over time. 


By considering the previous work, the study presented here is a new effort in the arena of social networks; one that by uses a large dataset of Facebook profiles to analyze the profile information disclosure patterns.

\section{Data Collection and Attributes Definition}
\label{sec:methodology}

We have developed an HTML crawler that is able to collect publicly-available information from a Facebook user's profile. The crawler collects up to 19 attributes from each profile. It must be noted that our tool respects the privacy of users since we only collect information that users themselves decide to share publicly. 
We run our crawler between March to June 2012 and captured the profile of 479k Facebook users randomly selected throughout the world. 
For each user we store up to 19 different attributes (only those publicly available). We classify those attributes into two categories: 
\\\textbf{Personal attributes:} Friend-list, Current City, Hometown, Gender, Birthday, Employers, College and HighSchool.
\\\textbf{Interest-based attributes}:  Music, Movie, Book, Television, Games, Team, Sports, Athletes, Activities, Interests and Inspired poeple.

The meaning of the personal attributes present are obvious and self-contained. It worth mentioning that some of them such as \textit{Employers, College,} or \textit{HighSchool} could include more than one item. 
We need to note that in our analysis we insert an ``artificial'' interest-based attribute, called \textit{Aggregate-Interests} which is a binary attribute, i.e. it is 1 if the user publicly shares at least one item among all the interest-based attributes, and 0 otherwise. The \textit{Aggregate-Interests} attribute lets us know if a user shares any interests without taking into account the separate categories.

Finally, in order to perform personal attribute correlations, and to gain further insights into some of them, we have divided our main dataset into several attribute-based groups. Basically, a given group $A$ includes all the users in our main dataset that publicly disclose attribute $A$. For instance, from this point onwards in the paper, when we mention the \textit{Gender} group we are referring to the group  that includes all the users in our dataset that make their gender available in their Facebook profile.


\section{Public exposure of Facebook profile attributes}
\label{sec:exposure_attributes}

In this section we define the degree of publicly disclosed information in Facebook
We first perform an attribute-based analysis to study the portion of Facebook users that disclose each attribute. 
Next, we study the correlation among pairs of personal attributes. 
This analysis will provide useful insights on whether some attributes are correlated and we will discuss some potential reasons for such correlation.  

\subsection{Degree of attributes disclosure}


\begin{table}[t]
	\centering
	\scriptsize
	\caption{Portion of users with publicly disclosed personal and interest-based attributes in Facebook profiles.}
	\scalebox{.9}{%
		\begin{tabular}{lrc}
			\toprule
			& \multicolumn{1}{c}{Attribute} &  \% Profiles accessible \\
			\midrule
			Personal & \multicolumn{1}{l}{Friend-list} & 62.7 \\
			
			attributes & \multicolumn{1}{l}{CurrentCity} & 36.1 \\
			& \multicolumn{1}{l}{Hometown} & 34.6 \\
			& \multicolumn{1}{l}{Gender} & 53.5 \\
			& \multicolumn{1}{l}{Birthday} & 2.9 \\
			& \multicolumn{1}{l}{Employers} & 22.5 \\
			& \multicolumn{1}{l}{College} & 16.8 \\
			& \multicolumn{1}{l}{HighSchool} & 13.2 \\
			\midrule
			Interest-based & \multicolumn{1}{l}{Aggregate-Interest} & 48.4 \\
			attributes    & \multicolumn{1}{l}{Music} & 41.0 \\
			& \multicolumn{1}{l}{Movie} & 28.3 \\
			& \multicolumn{1}{l}{Book} & 16.7 \\
			& \multicolumn{1}{l}{Television} & 31.8 \\
			& \multicolumn{1}{l}{Games} & 9.4 \\
			& \multicolumn{1}{l}{Team} & 8.5 \\
			& \multicolumn{1}{l}{Sports} & 2.3 \\
			& \multicolumn{1}{l}{Athletes} & 10.7 \\
			& \multicolumn{1}{l}{Activities} & 20.5 \\
			& \multicolumn{1}{l}{Interests} & 10.9 \\
			& \multicolumn{1}{l}{Inspire} & 1.9 \\
			\bottomrule
		\end{tabular}%
	}
	\label{tab:disclosed_all}%
\end{table}%

We provide some global numbers that paint a global picture of the amount of information (i.e. attributes) that Facebook users make publicly available. 
To this end first of all we study the default status of the attributes in Facebook. The study shows that out of the 479k analyzed users, only 11.62\% do not  share any attribute, 19.26\% disclose a single attribute, while the remanding users, 69.12\%, have two or more attributes in their profile that are publicly accessible. These values give a first reference point to  help understand that external users/entities can retrieve an enormous amount of information from Facebook profiles.

Our goal is to determine the level of privacy awareness that Facebook users present with respect to the different attributes. Table \ref{tab:disclosed_all} shows  the portion of users in our main dataset that publicly disclose each of the studied attributes. We first focus on personal attributes and then discuss interest-based attributes. 

\begin{table*}[t]
	\centering
	\scriptsize
	\caption{ Attributes correlation. Each value in the table refers to the portion of users belonging to the group indicated in the column that disclose the attribute indicated by the row.}
	\scalebox{.9}{%
		\begin{tabular}{rc|ccccccccc}
			\toprule
			\multicolumn{1}{c}{Attribute} & All  & Friend-list &  CurrentCity & Hometown &  Gender &  Age (Birthday) &  Job (Employers) &  College &  HighSchool \\
			\midrule
			\multicolumn{1}{l}{Friend-list} & 62.7 & 100  &  79.6 & 79.3 & 64.8 & 72.5 & 82.8 & 83 & 87 \\
			\multicolumn{1}{l}{CurrentCity} & 36.1 & 45.9 &  100  & \textbf{74}   & 42   & 56.2 & 55.4 & 59.4 & 57 \\
			\multicolumn{1}{l}{Hometown} & 34.6 & 43.7 &  \textbf{71} & 100  & 35.7 & 58.2 & 55.3 & 54.2 & 50.8 \\
			\multicolumn{1}{l}{Gender} & 53.5 & 55.3 &  61.7 & 55.2 & 100  & 58.8 & 55.7 & 79.9 & 86 \\
			\multicolumn{1}{l}{Birthday} & 2.9  & 3.4  &  4.6  & 4.9  & 3.2  & 100  & 5  & 4.9  & 4.2 \\
			\multicolumn{1}{l}{Employers} & 22.5 & 29.7 & 34.5 & 35.9 & 23.4 & 38   & 100  & \textbf{59}  & \textbf{53} \\
			\multicolumn{1}{l}{College} & 16.8 & 22.2  & 27.6 & 26.3 & 25 & 28 & \textbf{43.8} & 100  & 64.6 \\
			\multicolumn{1}{l}{HighSchool} & 13.2 & 18.3  & 20.8 & 19.3 & 21.2 & 18.7 & 31.1 & \textbf{50.7} & 100 \\
			\bottomrule
		\end{tabular}%
	}
	\label{tab:disclosed_O.ATT}%
\end{table*}%

\paragraph{Personal attributes}
The friend-list appears as the attribute with the greatest public exposure. Table \ref{tab:disclosed_all} shows that almost 63\% of the users make their friend-list available. This clearly indicates that FB users do not consider that exposing their connections could lead to any privacy issue. At the other extreme, the attribute with the lowest exposure is Birthday. Less than 3\% of the users reveal their age, which means that users regard this attribute as highly private. Here it worth to mention again that Birthday attribute is in the privacy level of ``only Friends" by default in Facebook and this 3\% of users they changed this level to publicly available.
Also, 1/2 of the users share their gender. A bit less, around 35\% of users, make their current city and their home town available publicly. This implies that users consider personal location information to be more sensitive than the information related to their contacts, but much less sensitive than their age. In addition, users seem to be more concerned about privacy issues linked to disclosing their job information since a little less than 1/4 of them publicly list their employers. We close the analysis of personal attributes by evaluating those related to education, where 17\% and 13\% of users publicly share their college and High School. Education-related attributes are thus the next-most private attributes after age. In summary, we can list the attributes in terms of public exposure (from more to less exposure) as follows: friend-list, gender, job, education, and age.

\paragraph{Interest-based attributes}
Table \ref{tab:disclosed_all} shows that almost 1/2 the users share at least one interest within the interests-based attributes, which means that Facebook users are not very concerned about the potential privacy implications that could be derived from sharing their interests. These attributes are initially less sensitive than personal attributes in terms of privacy. However, in some cases a particular interest of a user regarding some controversial issue could potentially lead to privacy issues. Looking at the results in the table we observe that the more popular categories are music (41\%), Television (32\%) and Movies (28.3\%). It is interesting that almost all users that share an interest (48\%) are actually sharing Music  (41\%). In contrast very few users share information in relation to their sports interest and as to what inspires them, just 2.3\% and 1.9\% respectively. The remaining interest-based attributes are made available by 10\%-20\% users. Finally, it is worth to mention that personal attributes such as Friendlist, CurrentCity, Hometown or Gender are more accessible than users' interests.

\subsection{Correlation of Facebook Attributes}

We now turn our attention to the different groups that include all the users that disclosed a particular attribute (e.g. CurrentCity), and how they correlate with the remaining personal attributes. Table \ref{tab:disclosed_O.ATT} shows the portion of users from a given group (columns) that share one of the remaining attributes (rows). For instance, the value crossing Current City column with Friend-list row means that 79.6\% of the users in the \textit{CurrentCity} group (i.e. those users from our dataset with their CurrentCity attribute available) also disclose their Friend-list. In addition, table \ref{tab:disclosed_O.ATT} includes the results obtained from our main dataset, referred to as \textit{All group} (the first column in the table), for comparison purposes.

First of all we observe that all the analyzed groups present a larger percentage for their available attributes than in \textit{All} group, which implies that users that share one personal attribute will likely share some other attributes. This assumption is supported by the observation that 2/3 of Facebook users disclose more than one attribute, as previously reported in this section.
It is especially noteworthy that most of the users (71\%) disclosing their \textit{CurrentCity} also make public their \textit{Hometown}, and close to 74\% of users that share their \textit{Hometown} attribute also disclose their \textit{CurrentCity}. This indicate that Facebook users relate these two attributes together, and in case they share the place where they currently live, they also disclose the place where they were born. In fact, these two parameters are the only ones that directly provides a physical location 
and it is clear that most Facebook users providing location information tend to share both of these indicators. Therefore, we can conclude that \textit{CurrentCity} and \textit{Hometown} attributes are highly correlated since 3/4 of users disclosing one of these attributes will also share the other one.

We also find a significant correlation when we relate the employment and the education attributes. The users composing the \textit{Employment} group tend to also share some educational information. In particular, 44\% of the users that make their job information available also show their College, and 31\% identify their High School. This is also validated in the other direction as 59\% and 53\% of users in the \textit{College} and \textit{HighSchool} groups, respectively, made their employer available. In addition, as we would expect, the two education attributes are highly correlated with each other. In contrast, user groups that are not related to education or employment information show a much lower correlation to these attributes, always below 38\%, 28\% and 21\% for Employment, College and HighSchool, respectively. 
Furthermore, the high number of users (44\%) disclosing their College within the \textit{Employers} is significant even though 44\% reflects less than half of all, that figure is quite high given that a large number of users in Facebook that cannot share their college because they simply never attended (or did not graduate). Then, that 44\% is actually a very relevant number that roughly demonstrates that whoever indicates their employer (or employment status) in Facebook and has obtained a University degree wants to make it public. This hypothesis is validated by the fact that only 31\% users in the \textit{Employers} group share its HighSchool, and obviously there are more users in the \textit{Employers} group who went to the High School than the ones who went to the University. Previous statement is validated by the fact that  65\% of users in \textit{HighSchool} group also report their College, whereas this portion is reduced to 50\% for those users in \textit{College} group that also report their \textit{HighSchool} information.  Therefore, we can extract two main conclusions from the correlation analysis between education and employment: $(i)$ These two attributes are clearly correlated in Facebook, and $(ii)$ an important fraction of users in Facebook understand that disclose the University they attended does not imply any privacy issue, instead they seem to believe it provides them with a good reputation.

In the \textit{Gender} group we do not find any strong correlations, only very weak correlations with \textit{CurrentCity} (42\%), \textit{Hometown} (36\%) and \textit{Employers} (23\%) compared to the correlations of the rest of the groups with these attributes. This would suggest that users sharing their gender have strong privacy concerns with respect to their location and employment information. 

In a nutshell, we have found strong correlations between: $(i)$ the \textit{CurrentCity} and \textit{Hometown} attributes, and $(ii)$ the education attributes, \textit{College and HighSchool}, between each other and with the \textit{Employment} attribute. We believe that the first correlation is because that users roughly perceive both parameters as location information, so if they do not have privacy concerns with one, they also do not have an issue with the other. Our hypothesis for the second correlation is that users, with anyone preparing a resume, find that education and employment attributes complement each other.



\section{Public Exposure of Facebook Users}
\label{sec:exposure_users}

To this point, we have performed an attribute-based analysis that has allowed us to understand which attributes are more privacy-sensitive for Facebook users, and to identify the correlation that exists (or not) among the different attributes. However, this analysis did not account for the public exposure of Facebook users. Towards this end we need to perform a user-based analysis. Instead of taking one attribute and counting how many users share it, we now need to look at individual users and determine how many attributes (among all those possible ones) she is disclosing. For that we take into account all 19 attributes 
(Personal + Interest-based attributes). We define a simple but functional metric named as Degree of Public Exposure (DPE), which ranges from 0 to 19. Basically, we go through the 19 parameters and whenever one can be accessed we sum +1 to the DPE value for that user. By defining this metric we are able to easily compare the level of profile's attribute openness without considering any kind of difference between the attributes.

\begin{table}[t]
	\scriptsize
	\centering
	\caption{Median and Mean of DPE metric}
	\scalebox{.9}{%
		\begin{tabular}{lcc}
			\toprule
			Attribute & Median & Mean \\
			\midrule
			All & 3    & 4.27 \\
			Friend-list & 5    & 5.61 \\
			Likes-list & 7    & 7.11 \\
			CurrentCity & 7    & 7.18 \\
			Hometown & 7    & 7.35 \\
			Gender & 4    & 5.26 \\
			Birthday & 7    & 7.60 \\
			Employers & 7    & 7.26 \\
			College & 8    & 7.95 \\
			HighSchool & 8    & 8.02 \\
			\bottomrule
		\end{tabular}%
	}
	\label{tab:median_whole}%
\end{table}%

\begin{figure}[t]
	\centering
	\includegraphics [width=0.38\textwidth,height=0.26\textwidth] {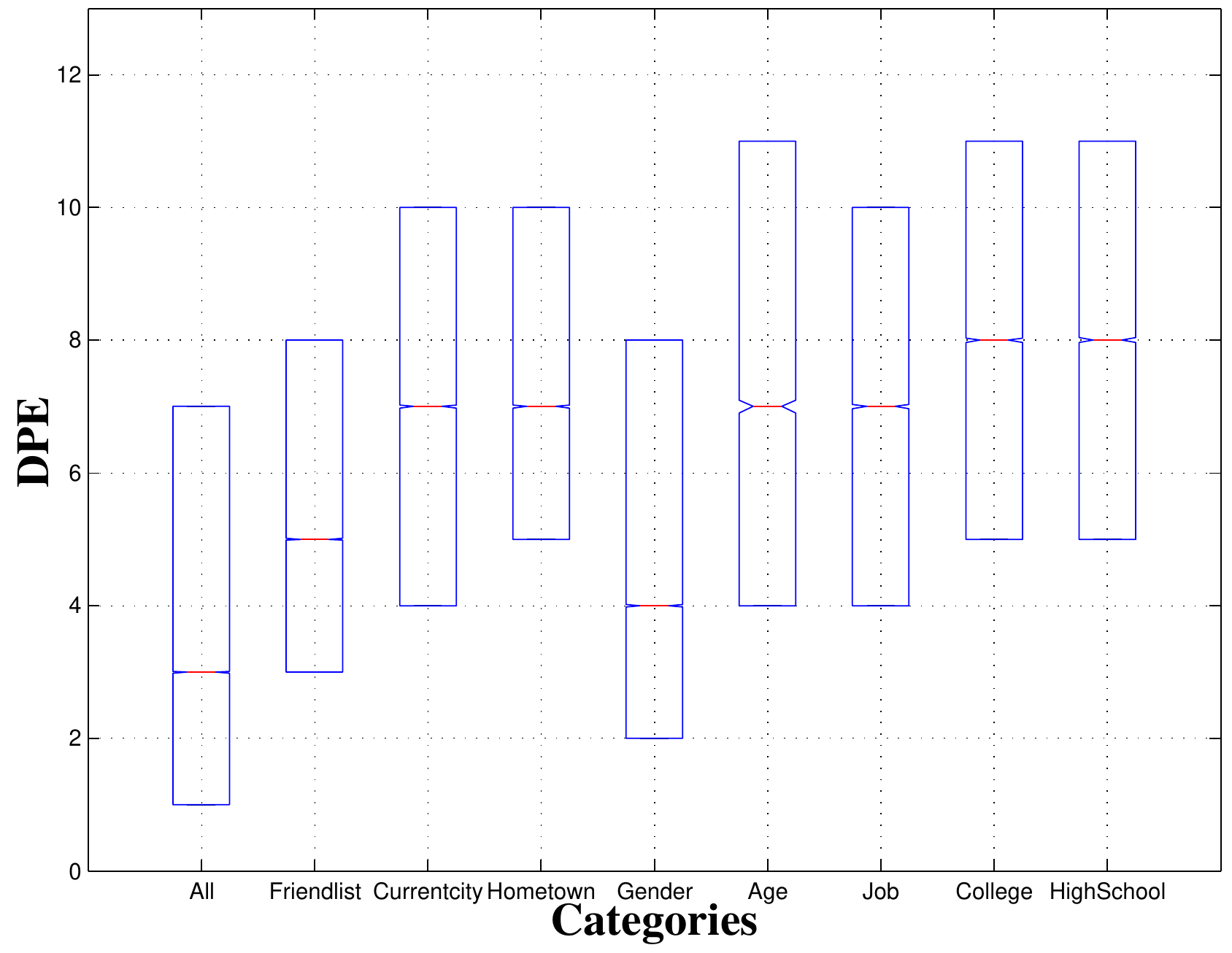}
	\caption{Box plot of DPE for categories}
	\label{fig:Metric_plotbox}
\end{figure}

Table \ref{tab:median_whole} shows the median and average value of the DPE metric for our main dataset, as well as each of the previous attribute-based groups, while Figure \ref{fig:Metric_plotbox} provides further details of the DPE distribution for the different groups by means of a box plot graph that shows the 25th, 50th (median) and 75th percentiles. If we first consider the results for \textit{All} group, we extract that a typical Facebook user presents an average DPE of 4.27. The remaining groups (except for \textit{Friend-list} and \textit{Gender}) show an average DPE higher than 7. This means that users in these groups publicly disclose more than seven attributes. It is worth noting that the users with a higher public exposure are those ones that share their education information, i.e  users in \textit{College} and \textit{HighSchool} groups, which present an average DPE of 7.95 and 8.02, respectively. If we analyze the results shown in Figure \ref{fig:Metric_plotbox}, we can observe that all the groups except \textit{All}, \textit{Friend-list} and \textit{Gender} present a DPE 75th percentile $\geq$10. This means that there are a relevant portion of users that disclose more than 10 attributes. Therefore, those users may be very attractive for external entities since they have a quite complete information regarding them.

In a nutshell, our results demonstrate that anyone can find substantial personal information from Facebook profiles since it is publicly available. In particular, our results suggest that if an entity wants to maximize the amount of information (i.e. attributes) retrieved from Facebook profiles, she should target users disclosing their education information.

\section{Examples of Public Facebook Information Usage}
\label{sec:Group_analysis}

In this section we show three examples of how the information available in Facebook can be used for different purposes. 

\subsection{Gender attribute: Men vs. Women public exposure}

In each attribute-based group we found users that provide their gender information and study which portion of them are males and which portion females. Table \ref{tab:Gender} shows the percentage of users for each gender and group. Male is the dominant gender for all the attributes except Birthday. This seems to indicate that generally men are less concerned about privacy issues than women, however the difference for most of the parameters is small, and never goes above 11 percentage points. The higher differences occur for Employers and HighSchool attributes. Finally, it is somewhat surprising that women share their age information slightly more frequently than men, which contradicts the ``cultural'' assumption that women tend to hide their age more often than men.

\begin{table}[ht]
	\centering
	\scriptsize
	\caption{Gender analysis per categories of attributes}
	\scalebox{.9}{%
		\begin{tabular}{rcc}
			\toprule
			\multicolumn{1}{c}{attributes' categories} & \%Male & \%Female \\
			\midrule
			\multicolumn{1}{l}{All} & 51.33 & 48.67  \\
			\multicolumn{1}{l}{Friend-list} & 53.99 & 46.01  \\
			\multicolumn{1}{l}{CurrentCity} & 52.81 & 47.19 \\
			\multicolumn{1}{l}{Hometown} & 54.05 & 45.95 \\
			\multicolumn{1}{l}{Gender} & 51.33 & 48.67  \\
			\multicolumn{1}{l}{Birthday} & 49.23 & 50.77\\
			\multicolumn{1}{l}{Employers} & 55.23 & 44.77  \\
			\multicolumn{1}{l}{College} & 53.30 & 46.70  \\
			\multicolumn{1}{l}{HighSchool} & 55.89 & 44.11 \\
			\bottomrule
		\end{tabular}%
	}
	\label{tab:Gender}%
\end{table}%

We can find many reports that explore gender differences in different disciplines like sociology and psychology such as \cite{Gender_Diff}, etc, which in many cases has a large diffusion even reaching general media. This example demonstrates that the publicly available information in Facebook is a potential source of information for these types of studies.

\subsection{Age distribution analysis}
\label{subsec:Age_analysis}

\begin{figure}[t]
	\centering
	\includegraphics [width=0.38\textwidth,height=0.24\textwidth] {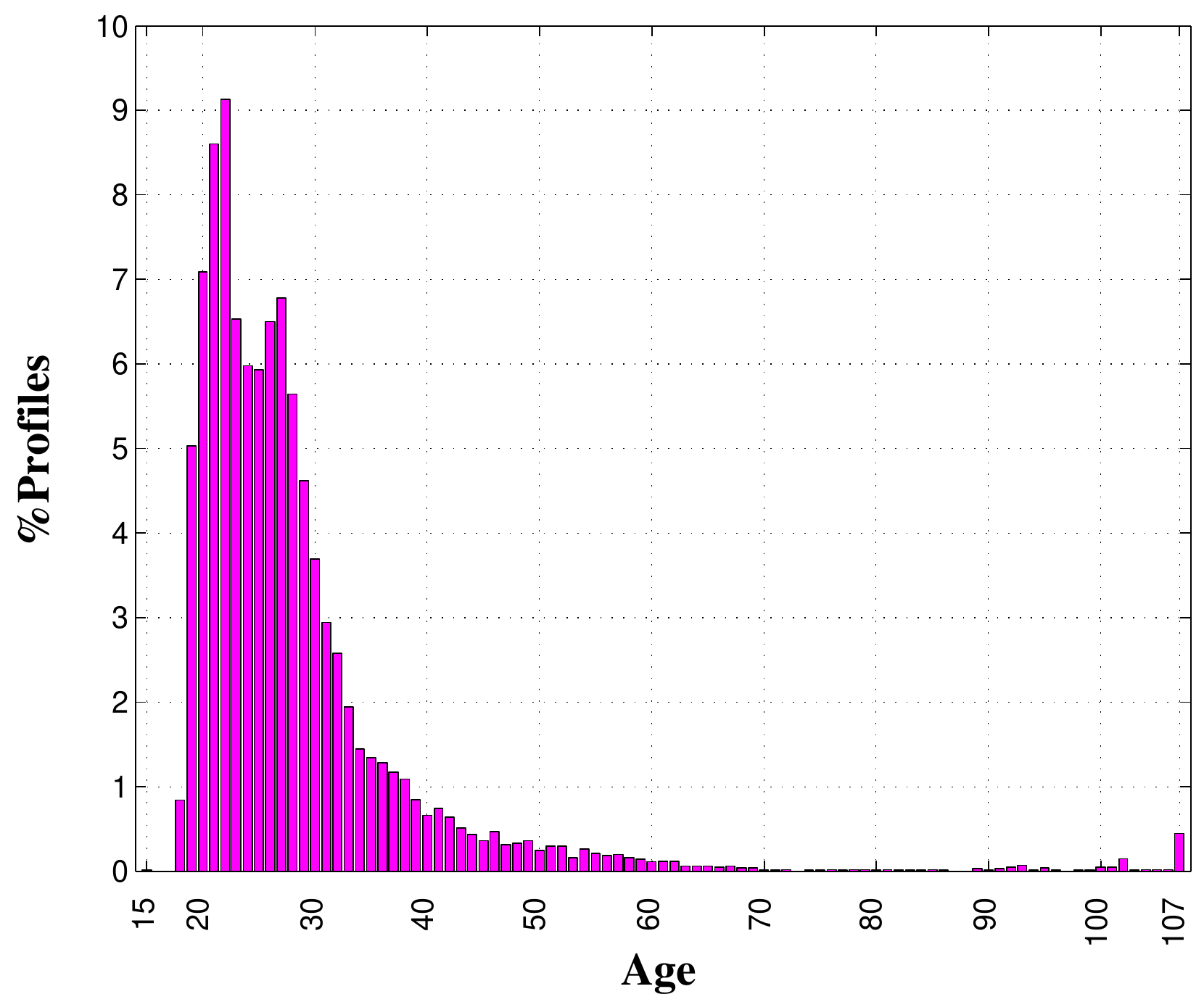}
	\caption{\%Profiles in different age range}
	\label{fig:age_profiles}
\end{figure}

%

Analyzing the distribution of ages among those few users (i.e. 2.9\% of the 479K) that publicly share their birth date reveals some unexpected results. Figure \ref{fig:age_profiles} shows the portion of users in our dataset belonging to each age from 13 to 107 (Facebook does not allow accounts to be opened for users younger than 13). Surprisingly, we found very few users $\leq$18 years old, and we did not find any Facebook rule that penalizes the disclosure of birthdays for users less than 18 years old. The ages in the interval of 19-28 contain more than 50\% of the users revealing their age, with 21 and 22 the most represented ages containing more than 8\% of the users each one. From 28 years upwards we found an exponential decrement, in some few cases reaching ages above 100. Particularly, we observe that almost 0.5\% users report an age of 107 (indicating 1905 as their birthday year, which at the time we collected the data was the oldest year allowed by Facebook). It is very likely that these are fake ages introduced by users who do not want to provide their real age.

\begin{table}[htbp]
	\centering
	\scriptsize
	\caption{Age of users with disclosed birthday and the gender distribution in different categories of age}
	\scalebox{.9}{%
		\begin{tabular}{lccc}
			\toprule
			Age category & \%Users in Birthday group & \%Female & \%Male \\
			\midrule
			Teenagers ($\leq$ 18) & 0,85 & 62,67 & 37,33 \\
			Post-Teenagers (19 - 25) & 48,29 & 54,83 & 45,17 \\
			Young (26 - 30) & 27,22 & 48,10 & 51,90 \\
			Mature (31 - 50) & 19,71 & 45,37 & 54,63 \\
			Senior & 3,93 & 37,54 & 62,46 \\
			\bottomrule
		\end{tabular}%
	}
	\label{tab:Age_User_Gender}%
\end{table}%

In order to provide aggregate numbers we have classified users into 5 different ages groups.
Table \ref{tab:Age_User_Gender} reports the portion of users included in each of these categories. The results  reveal that Teenagers very rarely (0.85\%) disclose publicly their age, which was totally unexpected statistic. In contrast, as the results confirmed our expectation that Senior users, which are the ones less representative in OSNs like Facebook, would present a low weight in the \textit{Birthday} group. Therefore, the big majority of users sharing their age belongs to he interval 18-50. In particular, Post-Teenagers between 18 and 25 years old represent 1/2 of the users sharing their birthday, followed by Young that accounts 1/4 of the users, and Mature group comprising 1/5 of the users from \textit{Birthday} group.

The results in the previous section revealed that women share their age a bit more often than men, and we want to check whether this is constant across different age categories. Table \ref{tab:Age_User_Gender} shows for each age category the portion of users whose gender is male or female. In the case of Teenage women expose their age much more than men. In the case of post-teenagers we find 10\% more women than men among the users disclosing their age. The observed tendency changes for young people between 26-30 years old where we find slightly more men sharing their age. This change of tendency is confirmed in  the Mature and Senior categories where there are 10\% and 25\% more men with open ages as compared to women, respectively. In summary, we can conclude that there is a clear trend, the younger the age group the larger the portion of women disclosing their age is as compared to men, and the other way around, the older the age group the more the portion of men disclosing their age.

As it happened for the gender analysis, there are other disciplines that use age groups to perform different types of analysis. We have demonstrated that Facebook allows researchers to easily identify users of particular ages who also have other personal and interest-based attributes accessible.

\subsection{CurrentCity population analysis}


In this section we aim to validate the accuracy of a small sample of Facebook to compute the distribution of worldwide population across cities according to their size. For this use case we need to perform a more complex analysis than in the previous use cases where the results were directly derived from our database.

We found 8,473 different cities in the \textit{CurrentCity} attribute inside our dataset. We used debepedia \cite{depepedia} (a crowdsourcing effort to extract structured information for Wikipedia) in order to retrieve the population associated to those cities. We were able to identify population for 1,840 cities that aggregately include 173,026 profile out of the users with open \textit{CurrentCity} attribute our database. We classify these cities into six categories according to their population. 
For each category we have extracted the portion of FB profiles (corresponding to those cities) belonging to each class. Furthermore, we've used official statistics reported by the United Nations (UN) in its 2011 World Urbanization Prospects report \cite{UN_City} (see page 25). Unfortunately, this report only includes granularity for cities with more than 500k citizens. Table \ref{tab:cities_population} collects the results for the FB and the UN report.

\begin{table}[t]
	\centering
	\scriptsize
	\caption{Population distribution of Facebook (\#Profiles) and world (\#Inhabitants) in different city size class}
	\scalebox{0.9}{%
		\begin{tabular}{ccr}
			\toprule
			City Size Class (\#Inhabitants) & \%Profiles (FB)  & \%Inhabitants (UN){\cite{UN_City}} \\
			\midrule
			$<$ 1K & 0.14 & \multicolumn{1}{c}{} \\
			1K - 10K & 3.60 & \multicolumn{1}{c}{} \\
			10K - 100K & 18.50 & \multicolumn{1}{c}{} \\
			100K - 500K & 18.39 & \multicolumn{1}{c}{50.9 ($<$500k)} \\
			500K - 1M & 8.78 & \multicolumn{1}{c}{10.10} \\
			1M -5M & 33.04 & \multicolumn{1}{c}{21.30} \\
			$>$ 5M & 17.55 & \multicolumn{1}{c}{17.07} \\
			\bottomrule
		\end{tabular}%
	}
	\label{tab:cities_population}%
\end{table}%

Facebook results reveal that less than 0.2\% of users live in small villages with less than 1K inhabitants. Our hypothesis for this result is that people living in such small villages is usually senior people ($>$50), which, as demonstrated in section \ref{subsec:Age_analysis}, is very low population in Facebook. Therefore, we believe this data may not reflect the reality. Larger villages up to 10K citizens are reported by 3.6\% of users. Again we think this data is biased by the same reason explained before. Small towns going from 10K to 100K citizens and big towns between 100k-500K inhabitants show the same portion of profiles, roughly 18.5\% each of them, so 37\% both categories together. We found almost 9\% of Facebook users in cities from 0.5M to 1M citizens. Finally, cities above 1M users include more than 1/2 users, which are divided as follows. One-third of Facebook users report that they live in cities with a population between 1M and 5M, and 17.5\% of the users live in very big cities with more than 5M. 

Here we compare the Facebook results to the UN data in order to check the accuracy of a small Facebook sample (i.e. users in our dataset belonging to those 1,840 cities for which we were able to identify their population) to estimate the worldwide population distribution across cities according to their size. First of all, our data is able to very accurately estimate the portion of worldwide population in cities with more than 5M citizens. Furthermore, we also found a quite accurate estimation of the population in cities whose population ranges between 500k and 1M, since there is a discrepancy a bit higher than 1\%. In contrast, we found an important discrepancy for the case of cities between 1M-5M citizens and towns whose population is less than 500k. In the former case our data assign 33\% of Facebook users to those cities, while UN data only reports 21\%, a 12 percentage point difference. This is aligned to the 11 percentage point difference for cities with less than 500k citizens, since our data predicts 40\% and UN data 51\%. We believe that part of this deviation is due to the small amount of users our data reports for villages below 10K users (less than 4\%), since probably this portion is considerably larger in reality, but we believe people on those villages shows a much lower penetration in the use of technology (including OSNs) and thus Facebook results are biased.


\section{Conclusion}
\label{sec:Conclusion}

In this paper with the goal of understanding the degree of Facebook profile's informartion disclosure, we study the privacy status of Facebook profiles by analyzing the profile's attributes disclosure degree in a dataset including 479K Facebook profiles publicly available information that we have crawled from March to June 2012. 
The analysis of this data reveals the following main insights about the disclosed information in Facebook profiles. \noindent $(i)$ Friend-list is the attribute with the largest public exposure, whereas Birthday attribute is the one showing major privacy concerns from Facebook users. \noindent $(ii)$ We find strong correlations between \textit{Current City} and \textit{Home Town} attributes as well as (i.e. \textit{College} and \textit{HighSchool}) and professional (i.e. \textit{Employers}) attributes. \noindent $(iii)$ In average Facebook users make more than 4 attributes publicly available in their profiles. \noindent $(iv)$ Men show a larger public exposure as compared to women for all personal attributes except \textit{birthday}. \noindent $(v)$ The more representative age range based on the public available information is 18-25 that accounts for 1/2 of the users among those ones making its Birthday publicly available. \noindent $(vi)$ We show that Facebook accurately estimates the portion of people leaving in different class of cities.

\section{Acknowledgments}
The research leading to these results was funded by the European
Union under the project eCOUSIN (EU-FP7-318398)
and the project TWIRL (ITEA2-Call 5-10029).



\begin{thebibliography}{10}
	\providecommand{\url}[1]{#1}
	\csname url@samestyle\endcsname
	\providecommand{\newblock}{\relax}
	\providecommand{\bibinfo}[2]{#2}
	\providecommand{\BIBentrySTDinterwordspacing}{\spaceskip=0pt\relax}
	\providecommand{\BIBentryALTinterwordstretchfactor}{4}
	\providecommand{\BIBentryALTinterwordspacing}{\spaceskip=\fontdimen2\font plus
		\BIBentryALTinterwordstretchfactor\fontdimen3\font minus
		\fontdimen4\font\relax}
	\providecommand{\BIBforeignlanguage}[2]{{%
			\expandafter\ifx\csname l@#1\endcsname\relax
			\typeout{** WARNING: IEEEtran.bst: No hyphenation pattern has been}%
			\typeout{** loaded for the language `#1'. Using the pattern for}%
			\typeout{** the default language instead.}%
			\else
			\language=\csname l@#1\endcsname
			\fi
			#2}}
	\providecommand{\BIBdecl}{\relax}
	\BIBdecl
	
	\bibitem{UN_City}
	\BIBentryALTinterwordspacing
	{United Nations}, ``World urbanization prospects, 2011 revision.'' [Online].
	Available: \url{http://esa.un.org/unup/pdf/WUP2011_Highlights.pdf}
	\BIBentrySTDinterwordspacing
	
	\bibitem{Quercia_Facebook_Privacy}
	D.~Quercia, D.~B.~L. Casas, J.~P. Pesce, D.~Stillwell, M.~Kosinski, V.~Almeida,
	and J.~Crowcroft, ``Facebook and privacy: The balancing act of personality,
	gender, and relationship currency,'' in \emph{ICWSM}, 2012.
	
	\bibitem{Quercia_popular_facebook}
	D.~Quercia, R.~Lambiotte, D.~Stillwell, M.~Kosinski, and J.~Crowcroft, ``The
	personality of popular facebook users,'' in \emph{CSCW}, 2012.
	
	\bibitem{privacy-facebook-gross}
	R.~Gross and A.~Acquisti, ``Information revelation and privacy in online social
	networks,'' in \emph{Proceedings of the 2005 ACM workshop on Privacy in the
		electronic society}, ser. WPES '05, 2005, pp. 71--80.
	
	\bibitem{ChangRBM10}
	J.~Chang, I.~Rosenn, and L.~Backstrom, ``epluribus: Ethnicity on social
	networks,'' in \emph{ICWSM}, 2010.
	
	\bibitem{Lewis_thetaste}
	K.~Lewis, J.~Kaufman, and N.~Christakis, ``The taste for privacy: An analysis
	of college student privacy settings in an online social network,'' in
	\emph{Journal of Computer-Mediated Communication}, 2008.
	
	\bibitem{All_about_me}
	A.~Nosko, E.~Wood, and S.~Molema, ``All about me: Disclosure in online social
	networking profiles: The case of facebook,'' \emph{Computers in Human
		Behavior}, vol.~26, no.~3, pp. 406--418, 2010.
	
	\bibitem{leakage_infocom}
	R.~Dey, C.~Tang, K.~Ross, and N.~Saxena, ``Estimating age privacy leakage in
	online social networks,'' in \emph{INFOCOM, IEEE}, march 2012.
	
	\bibitem{You_are_who_you_know}
	A.~Mislove, B.~Viswanath, K.~P. Gummadi, and P.~Druschel, ``You are who you
	know: inferring user profiles in online social networks,'' in \emph{WSDM
		'10}, 2010, pp. 251--260.
	
	\bibitem{lampe}
	N.~Lampe Cliff~A.C., Ellison and C.~Steinfield, ``A familiar face(book):
	profile elements as signals in an online social network,'' in \emph{SIGCHI,
		Conference on Human Factors in Computing Systems}, pp. 435--444.
	
	\bibitem{percom_Facebook}
	R.~Dey, Z.~Jelveh, and K.~W. Ross, ``Facebook users have become much more
	private: A large-scale study,'' in \emph{PerCom Workshops}, 2012.
	
	\bibitem{Gender_Diff}
	{Joiner R, Gavin J, Brosnan M, Cromby J, et al}, ``Gender, internet experience,
	internet identification, and internet anxiety: a ten-year followup,'' in
	\emph{Cyberpsychol Behav Soc Netw.}, 2012.
	
	\bibitem{depepedia}
	\BIBentryALTinterwordspacing
	{Depepedia Portal}, ``Depepedia: structured information of wikipedia.''
	[Online]. Available: \url{http://dbpedia.org/page/Paris}
	\BIBentrySTDinterwordspacing
	
\end{thebibliography}

\end{document}